\documentclass[preprint,12pt,authoryear]{elsarticle}

\usepackage{graphicx}%
\usepackage{multirow}%
\usepackage{amsmath,amssymb,amsfonts}%
\usepackage{amsthm}%
\usepackage{mathrsfs}%
\usepackage[title]{appendix}%
\usepackage{xcolor}%
\usepackage{textcomp}%
\usepackage{manyfoot}%
\usepackage{booktabs}%
\usepackage{algorithm}%
\usepackage{algorithmicx}%
\usepackage{algpseudocode}%
\usepackage{listings}%
\usepackage{pdfpages}
\usepackage[colorlinks=true, linkcolor=blue, citecolor=blue, urlcolor=blue]{hyperref}
\usepackage{booktabs}
\usepackage{threeparttable}

\begin{document}



\begin{frontmatter}



\title{Assessing the Impact and Underlying Pathways of Sequenced AI feedback on Student Learning} %% Article title

\author[1]{Jie Cao}
\ead{jiecao@unc.edu}

\author[2]{Chloe Qianhui Zhao}
\ead{cqzhao@cmu.edu}

\author[3]{Christian Schunn}
\ead{Schunn@pitt.edu}

\author[2]{Elizabeth A. McLaughlin}
\ead{mimim@andrew.cmu.edu}

\author[4,2]{Jionghao Lin}
\ead{jionghao@hku.hk}

\author[2]{Kenneth R. Koedinger}
\ead{koedinger@cmu.edu}

%% 机构统一集中声明
%% 前面的 \author 代码保持不变...

%% 使用 \address 替代 \affiliation，直接写成完整的字符串
\address[1]{The University of North Carolina at Chapel Hill, Chapel Hill, NC, USA}

\address[2]{Carnegie Mellon University, Pittsburgh, PA, USA}

\address[3]{The University of Pittsburgh, Pittsburgh, PA, USA}

\address[4]{The University of Hong Kong,Hong Kong, China}


\author{} %% Author name

%% Abstract
\begin{abstract}
Feedback is essential for learning, but its effectiveness relies heavily on how well it engages students in the educational process. Generative AI offers novel opportunities to efficiently produce rich, formative feedback, ranging from direct explanations to incrementally sequenced scaffolding designed to promote learner autonomy. Despite these capabilities, it is still unclear whether sequenced (layered) AI feedback—which provides encouragement and hints before revealing the correct answer—genuinely enhances engagement and learning outcomes. To investigate this, we randomly assigned 199 participants to receive either sequenced or non-sequenced AI-generated feedback. We evaluated its impact on learning performance, cognitive and behavioral engagement, and affective perceptions to understand how these factors mediate overall learning outcomes. Results show that sequenced feedback elicited slightly higher behavioral engagement and, as anticipated, was perceived as more encouraging and supportive of student independence. Concurrently, however, it induced a higher level of mental effort. Mediation analyses identified a positive affective pathway driven by perceived encouragement, which was completely counteracted by a negative behavioral pathway associated with the average number of tasks requiring three or more submissions; the cognitive pathway (mental effort) remained non-significant. Overall, sequenced feedback led to significantly poorer learning outcomes when compared to direct, non-sequenced feedback. These findings highlight a crucial trade-off: although sequenced AI scaffolding boosts engagement and positive user perceptions, it can have a detrimental effect on actual learning performance. By integrating analyses of outcomes, perceptions, and underlying mechanisms, this study provides nuanced insights for designing automated, AI-driven feedback systems.

\end{abstract}

\begin{highlights}

\item sequenced AI feedback led to poorer learning performance than non-sequenced AI feedback
\item It boosted perceived encouragement and independence, but raised mental effort.
\item It also enhances behavioral engagement slightly.
\item Encouragement aided learning, while frequent submission behaviors hindered it.
\item Design must optimize positive tone, limit gaming, and aid cognitive processing.

\end{highlights}

%% Keywords
\begin{keyword}
Sequenced AI Feedback \sep  Elaborated Feedback\sep Learner-centered Feedback\sep Behavioral Engagement\sep LLM-generated feedback


\end{keyword}

\end{frontmatter}

% abstract + intro (2pages)
\section{Introduction}
Feedback is widely recognized as critical for supporting knowledge acquisition, sustaining motivation, and promoting self-regulation \citep{Hattie2007}. As found in a recent meta-analysis, feedback exerts a moderate positive effect on learning outcomes across different educational levels and domains \citep{wisniewski2020power}. Yet, not all feedback is equally effective. Its impact depends on contextual factors such as the learning environment and learner characteristics \citep{heckler2016factors}, as well as intrinsic features of the feedback itself, including modality, delivery mode, source, and, most importantly, its content \citep{ramadan2024step,graham2015formative,10.1145/3706468.3706479}. 

A particularly effective form of feedback provides the correct answer with an explanation of why it is the correct answer \citep{heckler2016factors, van2015effects,butler2013explanation,wisniewski2020power}. Such elaborated feedback with the correct answer can deliver information directly and efficiently, and it can be further elaborated to include explanations, strategies, and guidance for self-regulation \citep{wisniewski2020power}, which shows effectiveness in producing larger learning gains, particularly for students with lower prior knowledge \citep{heckler2016factors}.

Nevertheless, important limitations remain. During the learning process, immediate provision of correct answers may encourage students to copy the answer without fully engaging in understanding \citep{kulhavy1977feedback}. Emerging learner-centered feedback theories emphasize that feedback should not merely transmit information, but rather foster active engagement, self-regulation, and shared responsibility in the feedback process \citep{Ryan18082021}. Effective feedback, from this perspective, with the feature of learner-centered, requires them to generate, process, and respond actively to information, while also receiving affective support such as encouragement \citep{carless2022teacher}. These insights suggest that withholding the correct answer initially, while offering encouragement and prompts for self-regulation, may promote deeper engagement.

\begin{figure}[h]
  \centering
  \includegraphics[width=1\linewidth]{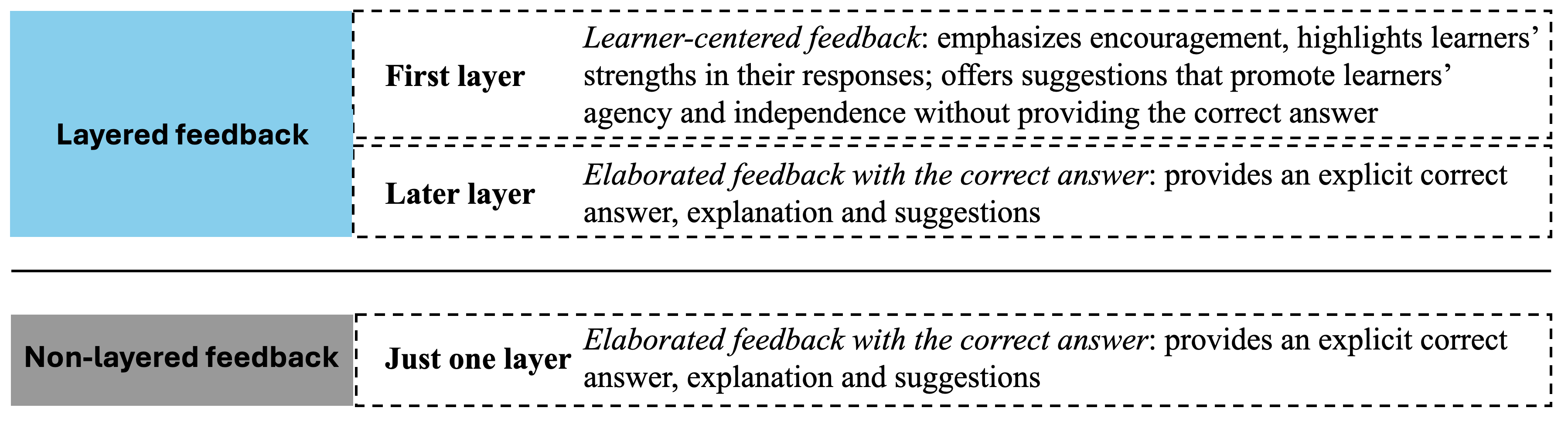}
  \caption{The design of the layered feedback compared with the non-layered feedback}
  \label{fig_layer feedback}
\end{figure}

Building on this reasoning, the present study introduces a dynamic \textbf{layered feedback} design (refers to the sequenced feedback) that integrates the strengths of learner-centered feedback and elaborated feedback with correct answers (See Figure \ref{fig_layer feedback} and details in \ref{Feedback conditions}). In this design, the first layer provides encouragement, promotes agency and independence, and includes actionable suggestions without revealing the correct answer, while a subsequent layer supplies elaborated, correct answer-based feedback if learners continue to struggle.  This approach to layering information is related to the sequential hint approach included in many Intelligent Tutoring Systems (ITS) systems that provide layer hints for students from basic pointing hints to final ``bottom-out hints'' (i.e., hints that provide answers, typically found at the end of a hint sequence) \citep{vanlehn2006behavior}. However, different from feedback, these hints in ITS are often provided at the student's request \citep{koedinger2006cognitive,aleven2016help} rather than depending up the submission of a solution for feedback. This distinction is important because some students ask for hints before any struggle or tend to request bottom-out hints while paying minimal attention to preceding hints, resulting in less learning \citep{mathews2008analysing,aleven2000limitations}. Layered feedback, therefore, has the potential to promote greater independence and deeper engagement during learning \citep{vanlehn2006behavior}.

At the same time, with the rapid development of online higher education and the surge in online learners, there has been a substantial increase in the demand for feedback. However, providing immediate feedback at scale is often infeasible for educators. Advances in large language models (LLMs) offer unprecedented opportunities to automate feedback generation that is rapid, adaptive, and multimodal \citep{caojie,10.1145/3706468.3706494,10.1145/3636555.3636901}. Further, the feedback can be easily elaborated to include personalized content. However, too much elaboration can be overwhelming. Generally, concerns have been voiced regarding AI providing direct answer information that guides task completion without requiring any learning \citep{kasneci2023chatgpt}.

 Prior studies have demonstrated both the feasibility and quality of LLM-generated feedback. However, existing work has typically relied on uniform prompts that produce structurally similar feedback throughout learning \citep{nguyen2024using,steiss2024comparing}. On the one hand, there is robust evidence that elaborate feedback with correct answers facilitates learning \citep{heckler2016factors, van2015effects, wisniewski2020power}. On the other hand, learner-centered feedback theory \citep{Ryan18082021}and studies on layered hints in ITS suggest that layered feedback holds significant potential for fostering independent and deep engagement \citep{vanlehn2006behavior}. Consequently, it remains to be determined whether layered LLM-generated feedback that combines learner-centered elements with delayed correct answers can yield superior learning outcomes compared to elaborated LLM-generated feedback.

 Furthermore, previous studies on LLM-generated feedback have primarily compared academic performance and user perceptions against human feedback (e.g., \citep{teng2024chatgpt,er_assessing_2024}). Consequently, there remains a lack of comparative analysis between different LLM-generated feedback, and the underlying mechanisms(how different feedback versions influence learning engagement, then finally influence final outcomes) remain unexplored. Thus, to address these gaps, we compared the effect of \textbf{LLM-generated layered feedback }and \textbf{non-layered feedback} on academic performance in an online higher education context, and investigated how learners' behavioral, cognitive engagement as well as affective perception differ in two feedback conditions and how do these differences lead to academic performance differences. Specifically, we propose four Research Questions (RQs): 

\begin{itemize}
    \item \textbf{RQ1:} Does layered feedback significantly improve learners' learning performance?
    \item \textbf{RQ2:} How do learners behaviorally engage with layered feedback compared to non-layered feedback?
    \item \textbf{RQ3:} How does layered feedback affect learners' perceptions (perceived specificity, encouragement, and independence) and their reported mental effort?
    \item \textbf{RQ4:} To what extent do the behavioral, affective, and cognitive factors mediate the effect of feedback type on learning performance?
\end{itemize}

% related work (1 to 2 pages)
\section{Related Work}
\subsection{Feedback in Education: Theoretical Foundations}
Feedback is a crucial element in teaching and learning, with the potential to shape students' learning outcomes, and is frequently employed to enhance their skills. Definitions of feedback vary, but one of the most frequently cited reviews on feedback in education described feedback as information provided by agents (e.g., teachers, peers, books) regarding progress or performance, guiding learners to go to the next step \citep{Hattie2007}.  This information enables feedback providers to clarify learning objectives (\textit{feed up}), illustrate learners' progress (\textit{feed back}), and offer guidance for closing the performance gap (\textit{feed forward}) \citep{Hattie2007}. However, different feedback designs can yield different influences (both positive and negative) on learners' motivation, autonomy, and learning outcomes \citep{wisniewski2020power,shute2008focus}.  Among them, the content or information embedded in feedback is a frequently-examined topic in educational research and practice \citep{ramadan2024step,graham2015formative}.

Research has shown that providing the correct answer is a crucial component of feedback, more useful than feedback merely indicating whether a learner’s response is correct or incorrect \citep{roper1977feedback,fazio2010receiving}. Feedback is even more effective when it includes both the correct answer and an explanation (often referred to as elaborated feedback or high-information feedback) \citep{wisniewski2020power} for learning and knowledge transfer \citep{butler2013explanation,swart2019supporting,van2015effects,heckler2016factors}. Further, \citet{heckler2016factors} found that the benefits of elaborated feedback were particularly pronounced for learners with lower prior knowledge and lower course grades. These conclusions are further supported by meta-analytic evidence: a synthesis of 40 studies comprising 70 effect sizes revealed that elaborated feedback (e.g., providing explanations) produced a substantially larger mean effect size than feedback limited to correctness judgments or feedback providing only the correct answer \citep{van2015effects}.

However, feedback that immediately supplies learners with correct answers can be limiting for many learners. Once answers are revealed, learners often disengage from deep thinking and simply copy the solutions \citep{kulhavy1977feedback}, a tendency that is especially detrimental to learners' agency in problem-solving contexts. Moving beyond the mere transmission of information, recent scholarship has conceptualized feedback as a learning process rather than just provided information (i.e., correct answers and explanations). Synthesizing insights from 95 studies, \citet{Ryan18082021} proposed a learner-centered feedback framework involving 12 key design attributes organized into three layers: contexts, characteristics, and components. Key characteristics include encouraging positive affect, inviting dialogue, and fostering independence, and key components highlight providing actionable suggestions for improvement and encouragement of learner agency. Together, these attributes reframe feedback as a developmental and dialogic process that empowers learners with agency and evaluative judgment, moving beyond transmission-focused models toward sustainable, learner-driven feedback practices. In sum, these learner-centered perspectives emphasize student agency, independence, and positive affect \citep{carless2022teacher}.

\subsection{Proposed layered feedback}
Given that both elaborated feedback with correct answers and learner-centered feedback have demonstrated distinct advantages, we propose the concept of \textit{layered feedback}, defined as a structured two-step approach in which feedback is delivered in two distinct layers with different pedagogical purposes (see details in the Research Design section \ref{Feedback conditions}).  

The design of layered feedback shares similarities with the design of hints in ITS. According to the standard hints design principle in ITS \citep{vanlehn2006behavior}, the hint sequence consists of three levels of increasing specificity: 1) Pointing Hint: Directs the student's attention to a specific error or location without revealing the solution \citep{hume1996hinting}. 2) Teaching Hint: Provides the domain principle or rule required to solve the step \citep{hume1996hinting}. 3) Bottom-Out Hint: Explicitly states the answer to the step to allow the student to proceed \citep{aleven2000limitations,vanlehn2006behavior}.  

Similarly, our proposed layered feedback design prioritizes initial guidance, providing a ``Bottom-Out Hint'' that explicitly reveals the solution only in the later layer. While hints are generally effective, some students tend to immediately request bottom-out hints, resulting in less learning \citep{mathews2008analysing,aleven2000limitations}. ITS hints are requested primarily by students on demand, whereas feedback is provided after task completion. Thus, it is possible that layered feedback will be more effective because students cannot bottom out without doing initial work. Overall, learner-centered theory and prior research on layered hints suggest that layered feedback may be superior to non-layered feedback. However, non-layered feedback can be effective without layering (e.g., \citep{heckler2016factors,van2015effects,wisniewski2020power}).  Consequently, the evidence on the relative effectiveness of these feedback structures remains inconclusive. By comparing layered feedback with non-layered feedback, we aim to generate new insights and empirical evidence into how feedback design influences both learning processes and outcomes, thereby informing educational practice.

\subsection{LLM-generated feedback}
Large Language Models (LLMs) have demonstrated substantial capabilities in educational practice, e.g., automatic grading \citep{Ferreira_Mello}, answer generation \citep{Rodrigues}, and student-chatbot interaction \citep{Woollaston, hao_mapping_2026,zhang-etal-2025-simulating}.  In addition, the natural language understanding and generation abilities of LLMs enable them to provide detailed, fluent, and coherent feedback that can surpass human instructors in some effective feedback attributes \citep{dai_assessing_2024}.  A growing body of research has examined the performance of LLM-generated feedback, which can be broadly grouped into three dimensions: evaluations of generated feedback quality \citep{caojie}, effects on learning outcomes\citep{10.1145/3636555.3636901}, and learners' perceptions \citep{10.1145/3706468.3706488}. With respect to evaluations of feedback quality, studies have focused on the correctness of generated feedback, the alignment with human feedback \citep{gabbay_combining_2024}, or alignment with educational theories \citep{dai_assessing_2024, caojie}. Regarding effects on learning outcomes, several studies report that LLM-generated feedback can yield comparable or even superior academic performance relative to human feedback \citep{nguyen_comparing_2024,ouyang_comparing_2024}, despite some negative findings \citep{er_assessing_2024}. Finally, regarding learners' perceptions of LLM-generated feedback, some studies indicate that students perceive such feedback as useful in identifying errors and offering constructive guidance \citep{ouyang_comparing_2024,zhaoslideitright}, but concerns about fairness and trustworthiness remain \citep{er_assessing_2024,zhaoslideitright}.  However, despite this growing literature, most studies employ a single prompt pattern to generate structurally similar feedback within their contexts, i.e., \citep{zhaoslideitright, nguyen_comparing_2024}. Research has yet to explore dynamic, layered approaches, leaving open the question of how LLM-generated layered feedback, compared to LLM-generated non-layered feedback, influences learning processes and outcomes.

\subsection{The mechanisms of feedback on learning outcomes}
The impact of feedback on learning outcomes is mediated through complex internal mechanisms. Three commonly-discussed mechanisms involve different dimensions of learners' engagement \citep{fredricks2004school}: cognitive processing, behavioral patterns, and affective states\citep{schiller2024understanding}. 

(a) \textit{Changing Cognitive Engagement}: Cognitively, the primary function of feedback is to bridge the gap between current understanding and the desired learning goal \citep{Hattie2007}. However, the mere provision of information is insufficient; effective feedback needs to trigger deep cognitive processing. As indicated in the ICAP framework (Interactive, Constructive, Active, Passive) \citep{chi2014icap}, learning outcomes are maximized when students construct knowledge from feedback rather than passively receiving it or mechanically implementing suggested changes. This process requires the investment of mental effort to interpret the feedback, diagnose errors, and adapt cognitive strategies \citep{wofford1990effects} after receiving feedback. Layering feedback may therefore improve learning by requiring learners to do more cognitive work before receiving the correct answers but also at the risks of overwhelm them.

(b) \textit{Changing Behavioral Engagement}: Effective feedback may also prompt productive regulation behaviors, such as revising answers, seeking help, or generally increasing time-on-task \citep{zimmerman2002becoming}, which increases learning opportunities and can result in positive effects on learning outcomes \citep{schiller2024understanding}.  Layered feedback can encourage more behavioral engagement simply by virtue of requiring extra submissions before receiving the correct answer feedback. However, not all triggered behaviors are effective for learning \citep{zhao2026llm} , and sometimes there are some meaningless and even harmful behaviors, e.g., ``gaming the system'' \citep{baker2008students}. 

(c) \textit{Changing Affective Perception}: From an affective perspective, feedback serves as a powerful modulator of learners' emotions and motivation, whether positive or negative \citep{belschak2009consequences}. According to Control-Value Theory (CVT) \citep{pekrun2006control}, feedback is a critical environmental factor that conveys information regarding controllability and academic value. When feedback provides clear guidance and supports a learner’s sense of competence, it acts as a positive antecedent that bolsters their appraisals of control and value. Layered feedback have the potential to offer opportunities to motivate students by encouraging tone or enhancing student's sense of independence without prematurely giving full solutions \citep{Ryan18082021}. Conversely, feedback that is vague or overly critical may induce anxiety and reduce self-efficacy \citep{shute2008focus}.  

In summary, layering feedback via LLMs, while not previously investigated, has the potential for improving learning outcomes cognitive, behavioral, and effective pathways, although negative effects are also possible. In the reported study, we examined learning outcomes as well as separate effects on each of the pathways.

% system + method  (3 to 5 pages)
\section{Method}
\subsection{Participants}
This study was approved by the [anonymized] University's Institutional Review Board.  Prolific\footnote{https://www.prolific.com/} was used to recruit a diverse pool of college students across various universities and academic backgrounds, ensuring a sample that is representative of university online learners. They received $\$15$ compensation and were informed of their right to withdraw at any time. A total of $215$ U.S. college students (109 female; $M_{\text{age}} = 32$) were recruited, and all of them provided pre-test, post-test, and process data. The participants represented diverse academic majors, with computer science and biology being the most prevalent. Participants were randomly assigned to either a layered ($n = 105$) or non-layered ($n = 110$) feedback condition. However, following screening of attention checks in the surveys, the final dataset included $199$ valid participants ($100$ from the layered group and $99$ from the non-layered group).

\subsection{Research design}\label{Research design}
\subsubsection{Learning content and learning-by-doing activities}
The learning content was drawn from the \textit{Multimedia Principle} chapter within a graduate-level book on E-learning principles (e-Learning and the Science of Instruction: Proven Guidelines for Consumers and Designers of Multimedia Learning) . Two learning goals from the chapter were selected: \textit{Describe the multimedia principle and why it is effective for learning}; and \textit{Recognize when the multimedia principle has been violated and when it has been applied well}. 

We then collaborated with a professor teaching this content to design $13$ learning-by-doing tasks, including eight Multiple Choice Questions (MCQs) and five open-ended questions. Upon submitting a response to each learning-by-doing task, learners received feedback that they could use to revise and resubmit their answers. This cycle continued until they arrived at the correct solution or completed the study.

\subsubsection{Feedback conditions} \label{Feedback conditions}
During the learning-by-doing activities, learners are randomly assigned to condition:

\textbf{Layered feedback condition (experiment group):} learners received layered feedback generated by the feedback system after submitting their response. In the first layer, feedback followed a learner-centered framework: it emphasized encouragement, highlighted learners' strengths, and promoted learners' independence without providing the correct answer, but provided usable hints and advice (learner-centered feedback) \citep{Ryan18082021}. This feedback layer aimed to maintain positive emotions while also guiding learners in self-correction. In the second layer, if the learner continued to struggle or fail in the tasks, the system provided an explicit correct answer, explanation of the correct answer, and suggestions (elaborated feedback with the correct answer), which aimedto directly show the information to students with full scaffolding. It has been found to be effective for learning in a meta-analysis \citep{wisniewski2020power}. Note that if students' responses are correct in the first attempt, they still receive learner-centered feedback highlighting their strengths; however, no suggestions of correct answer will be provided since they already know that but contain some advice for further learning.

\textbf{Non-layered feedback condition (control group):} learners in this condition also received feedback generated by the feedback system after submitting task responses. However, this feedback was not layered; students received feedback in the same structure each time, consisting of the explicit correct answer, explanation of the correct answer, and suggestions. Since the feedback is produced by an LLM, it is similar to but not necessarily identical to what the experimental group received as the second layer. In addition, the feedback remains identical regardless of the correctness of the student's response. If a student answers correctly, the system first confirms the accuracy and then provides an explanation of why the choice is correct to reinforce their understanding. The key distinction between the experimental group and the control group lies in the presence of the first layer of learner-centered feedback in the experimental group (See Figure \ref{fig_layer feedback}).  

During the learning-by-doing process, learners were encouraged to arrive at the correct answer or to attempt each question at least three times  (i.e., before feedback, after layer 1, after layer 2) to make full use of the feedback provided by the system.

\subsubsection{Feedback generated by LLMs}
\begin{figure}[ht!]
  \centering
  \includegraphics[width=\linewidth]{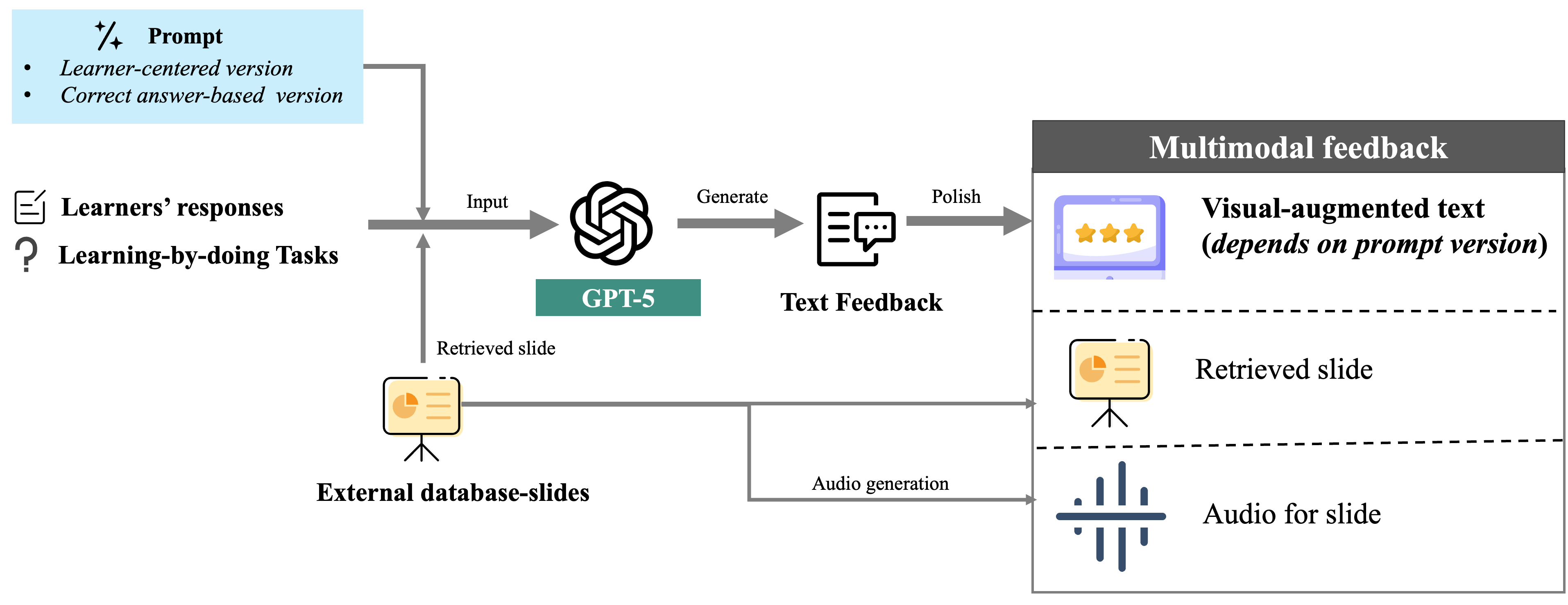}
  \caption{Feedback generation process}
  \label{fig_feedback generation}
\end{figure}

Both layered feedback and non-layered feedback (including learner-centered feedback and elaborated feedback with correct answers) were generated through the same process, differing only in the prompts applied. The overall workflow is illustrated in Figure \ref{fig_feedback generation}. First, the learner’s response and the corresponding learning-by-doing tasks were used as input data, and combined with one of two different prompts: a learner-centered feedback prompt and a correct answer-based feedback prompt. Both prompts contained sufficient background information, but they incorporated different rule-based constraints grounded in educational theory (see full prompts in the appendix). 

At the same time, we constructed an external database from instructional slides used by teachers and adopted a retrieval-augmented generation (RAG) approach to enhance feedback accuracy. The LLM model employed in this study was GPT-5 (\texttt{gpt-5-2025-08-07}). 

In both conditions, the text-based feedback generated via RAG was then formatted with visual augmentations (e.g., highlights and annotations) to improve engagement \citep{zhaoslideitright}. The retrieved slides were displayed alongside the feedback, and a speech module based on \texttt{gpt-realtime} generated audio narration of the slides, allowing learners to play and listen to the narrated content. 

\begin{figure}[ht!]
  \centering
  \includegraphics[width=\linewidth]{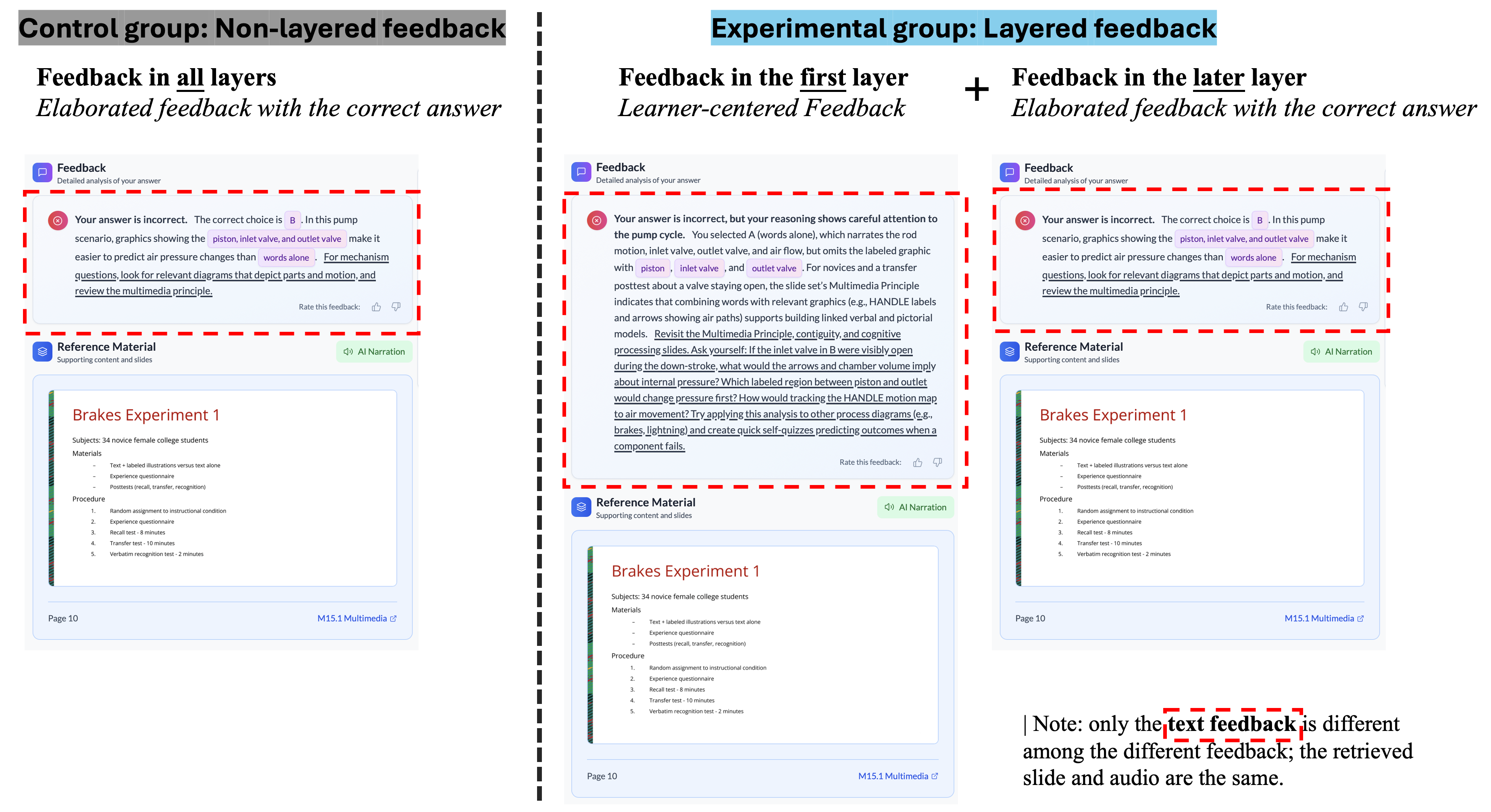}
  \caption{Generated feedback example in experimental and control groups}
  \label{fig_feedback example}
\end{figure}

Examples of generated feedback are presented in Figure \ref{fig_feedback example}. The left side presents an elaborated feedback with correct answer example from the control group (and with retrieved slides and audio). The right side presents an example of layered feedback from the experiment group (and also with retrieved slides and audio). Note that while the LLM-generated feedback comprises text, slides, and audio, the experimental manipulation only involves the textual components that were generated via different prompts. 

\subsection{Measurements}
\textit{Perceived Specificity ($\alpha$ = .76)} is the degree to which learners perceive feedback as specific and containing sufficient details. It was measured based on a scale developed by \citet{Zhang_Dong_Shi_Price_Matsuda_Xu_2024} and involved three items (e.g.,, \textit{``The feedback I received in the first try is specific and clear''}) on a five-point Likert scale (1 = strongly disagree, 5 = strongly agree).

\textit{Perceived Encouragement ($\alpha$ = .88)} refers to the extent to which learners perceived themselves as being encouraged and positive after receiving feedback, and represented support for affective engagement. It was measured using the 3-item scale developed by \citet{STRIJBOS2010291} (e.g., \textit{``The feedback I received acknowledged my good points or ideas''}) and employed a five-point Likert scale (1 = strongly disagree, 5 = strongly agree).

\textit{Perceived Independence ($\alpha$ = .75)} refers to the extent to which learners felt that they had agency and could finish tasks independently after receiving feedback. Based on the learner-centered feedback framework by \citet{Ryan18082021}, we developed a three-item survey to measure students’ perceived independence (e.g., \textit{``The feedback I received prompted me to have more learning agency''}). Responses were rated on a five-point Likert scale (1 = strongly disagree, 5 = strongly agree). 

\textit{Mental effort} refers to the perceived mental effort involved in processing the feedback, a form of cognitive engagement. It was measured via a single item \textit{``Please rate the mental effort you invested in understanding the feedback in this system''} on a 9-point mental effort rating scale originally developed by \citet{paas1992training} (1 = very, very low mental effort; 9 = very, very high mental effort). 

\textit{Behavioral engagement} refers to the extent to which students interacted with the learning-by-doing activities. It was measured using several indicators derived from the learning log data as developed by \citet{saini2019implementing}. Key metrics included learners’ overall time on task, the total number of submissions per learner, the number of tasks involving three or more submissions, and the edit distance for open-ended questions. The third metric served as a proxy for distinguishing the interaction patterns between the two groups: while the non-layered feedback group typically followed a two-step pattern (initial submission and another submission followed by receiving the answer), the layered feedback group engaged in a three-step iterative process (initial submission, revision based on the first layer of feedback, and a final submission based on subsequent feedback layers). In addition, the edit distance for open-ended questions were included to quantify learners' answer revision behavior.

\textit{Pre and Post test} Pre- and post-tests were co-designed with course instructors to assess learners’ mastery of the multimedia principle. The test contained 16 items (20 points total), including 14 multiple-choice (one point for each) and 2 open-ended questions (one has two points, one has four points).  Multiple-choice questions were automatically scored by the system. Open-ended responses were evaluated to give a score using an LLM-as-a-judge approach (GPT-4o). To assess the reliability and validity of this LLM-as-a-judge scoring method, we randomly sampled 10\% of the data for manual evaluation by human raters. The correlations between the LLM and human scores for the two questions were $r = .86$ and $r = .94$, respectively, indicating a high level of scoring consistency. 

\subsection{Research procedure}
\begin{figure}[h]
  \centering
  \includegraphics[width=0.9\linewidth]{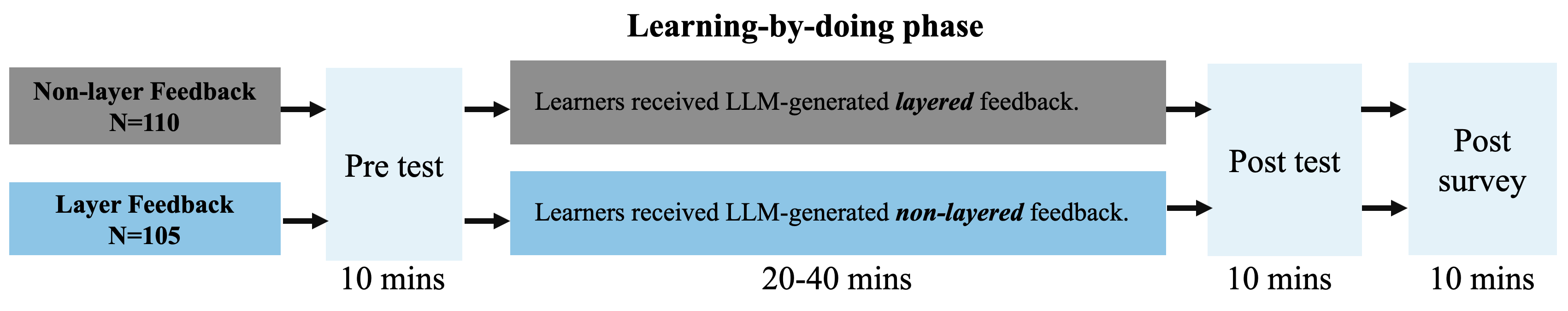}
  \caption{The research procedure}
  \label{fig_process}
\end{figure}

The overall research procedure is presented in Figure \ref{fig_process}. First, learners completed an informed consent form and were then randomly assigned to one of two groups. Next, they took a pre-test (10 minutes) before entering the learning-by-doing phase, during which they practiced independently and received different types of feedback. The experimental group received layered feedback; the control group received non-layered feedback. The learning-by-doing phase lasted around 20-40 minutes, depending on the learners' learning process. Afterward, learners completed a post-test, which was identical to the pre-test. Finally, they filled out a self-report questionnaire on their learning experience and perception (including mental effort, perceived specificity, perceived encouragement, perceived independence), along with demographic information.

\subsection{Data collection and analysis}
For RQ1, an Independent sample T-test was conducted to test whether there is a pre-performance difference between two groups. Then Analysis of Covariance (ANCOVA) was conducted on post-test performance as a function of experimental vs. control group, including learners' pre-test performance as a covariate.

For RQ2, we analyzed students’ behavioral engagement with the learning-by-doing activities. 
For these indicators, we conducted descriptive analyses and inferential tests (independent-samples t-tests and Mann-Whitney U test according to the distribution) to identify differences in behavioral engagement between groups.

For RQ3, we conducted independent-sample t-tests on the four survey-based scales (three perception dimensions and mental effort) to compare differences between the experimental and control groups. In addition, paired-samples t-tests were conducted within the experimental group to compare perceptions of learner-centered feedback in the first layer versus elaborated feedback with the correct answer in the latter layer.

To answer RQ4, we performed a parallel mediation analysis using Mplus. The model examined how the layered feedback (compared to non-layered feedback) influenced learning performance (Post-test) through four potential mediators that showed condition differences: perceived encouragement, perceived independence, mental effort, and the number of tasks requiring three or more submissions. We also included Pre-test scores as a covariate to control for learners' prior knowledge. We used bootstrapping with 5,000 resamples to test the significance of the mediation effects. This method provides robust estimates for normal and non-normal data. We determined statistical significance based on 95\% bias-corrected confidence intervals (CI); if the CI range did not include zero, the indirect effect was considered significant.

% results (3 to 4pages)
\section{Results}
\subsection{Non-layered feedback led to significantly greater learning performance}
Addressing RQ1,  we first compared the pre-test scores between the non-layered and layered feedback groups. The analysis revealed no significant differences ($p > .05$), confirming that both groups were comparable at baseline. Subsequently, an ANCOVA was conducted (see Table \ref{tab:ancova_combined}), with pre-test performance treated as the covariate. The model was statistically significant, $F(2, 196) = 50.39, p < .001$, accounting for 34\% of the variance ($R^2 = .34$). After controlling for initial performance, students in the non-layered feedback group achieved significantly higher post-test scores ($M_{adj} = 11.24, SE = 0.25$) than those in the layered feedback group ($M_{adj} = 10.40, SE = 0.25$). Contrast analysis further substantiated this effect, $t(196) = 2.32, p = .02$. These findings demonstrate that non-layered feedback was more effective in enhancing learning than layered feedback.

\begin{table}[htbp]
\centering
\footnotesize
\caption{ANCOVA Results for Post-test Performance by Feedback Condition after Controlling Pre-test}
\label{tab:ancova_combined}
\begin{tabular}{lccccccc}
\toprule
Group & $N$ & Mean & $SD$ & Adjusted Mean & Adjusted $SE$ & $F$ & $\eta^2_p$ \\
\midrule
Non-layered & 99  & 11.23 & 3.00& 11.24 & 0.25 & 5.39\textsuperscript{a} & 0.027\\
Layered     & 100 & 10.41 & 3.17& 10.40 & 0.25 & & \\
\bottomrule
\multicolumn{8}{l}{\textsuperscript{a} $p < 0.05$.}
\end{tabular}
\end{table}

\subsection{The complex behavioral engagement across two feedback versions}
To answer RQ2, we analyzed behavioral data during the learning-by-doing phases. 

\begin{figure}[h]
  \centering
  \includegraphics[trim={0 1cm 0 1cm}, clip,width=0.5\linewidth]{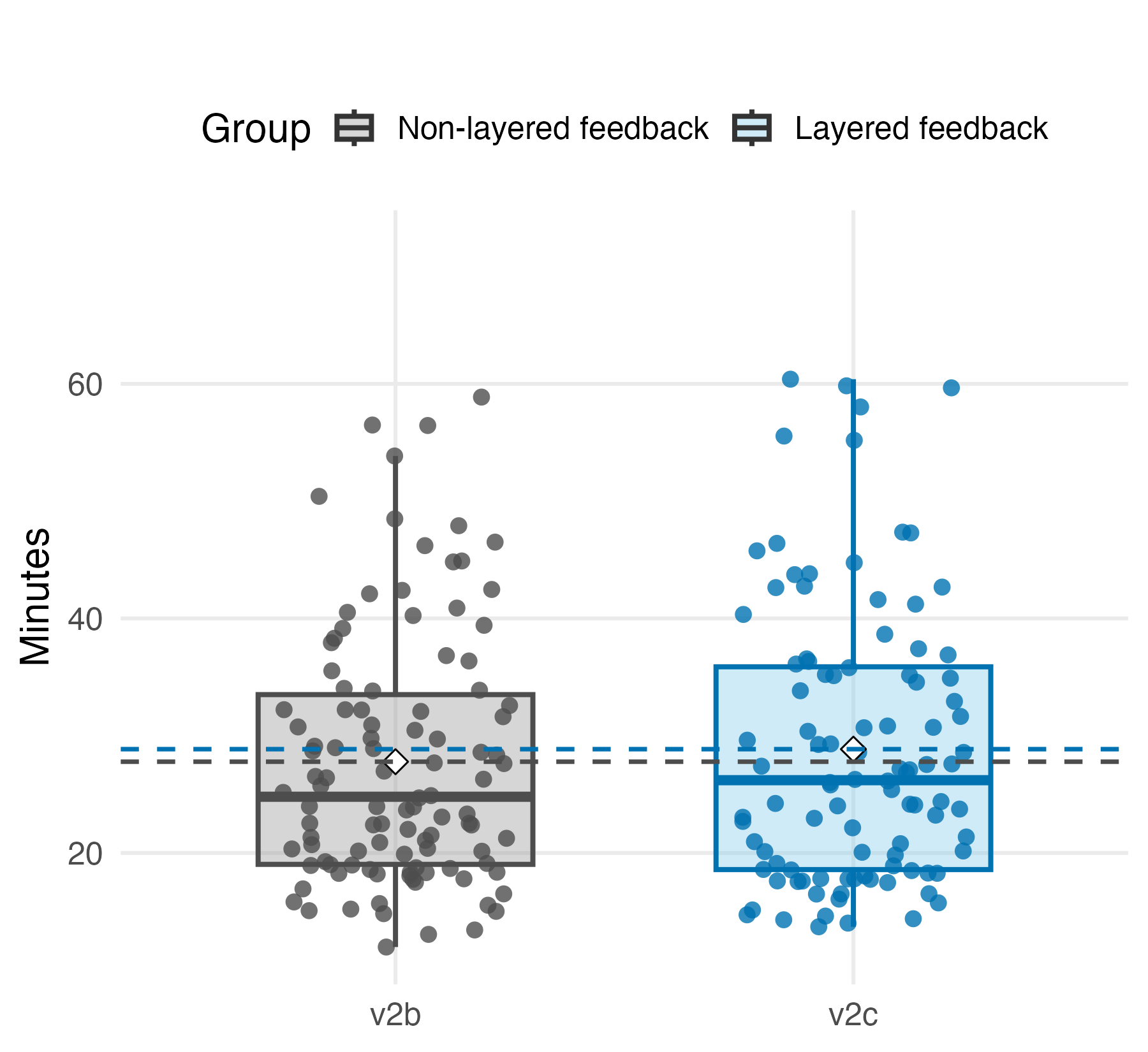}
  \caption{Overall learning time of both groups}
  \label{fig4}
\end{figure}

\textbf{(a). Overall learning time.}
We first compared the total learning time for the learning-by-doing phase. As shown in Figure \ref{fig4}, learners in the non-layered feedback group spent on average $29.3$ minutes ($SD = 13.5$), while learners in the layered feedback group spent a little bit more time, with $31.1$ minutes ($SD = 15.6$). The difference in time-on-task was not statistically significant ($p=.38$).

\begin{figure}[h]
  \centering
  \includegraphics[trim={0cm 0cm 0 0cm}, clip,width=0.8\linewidth]{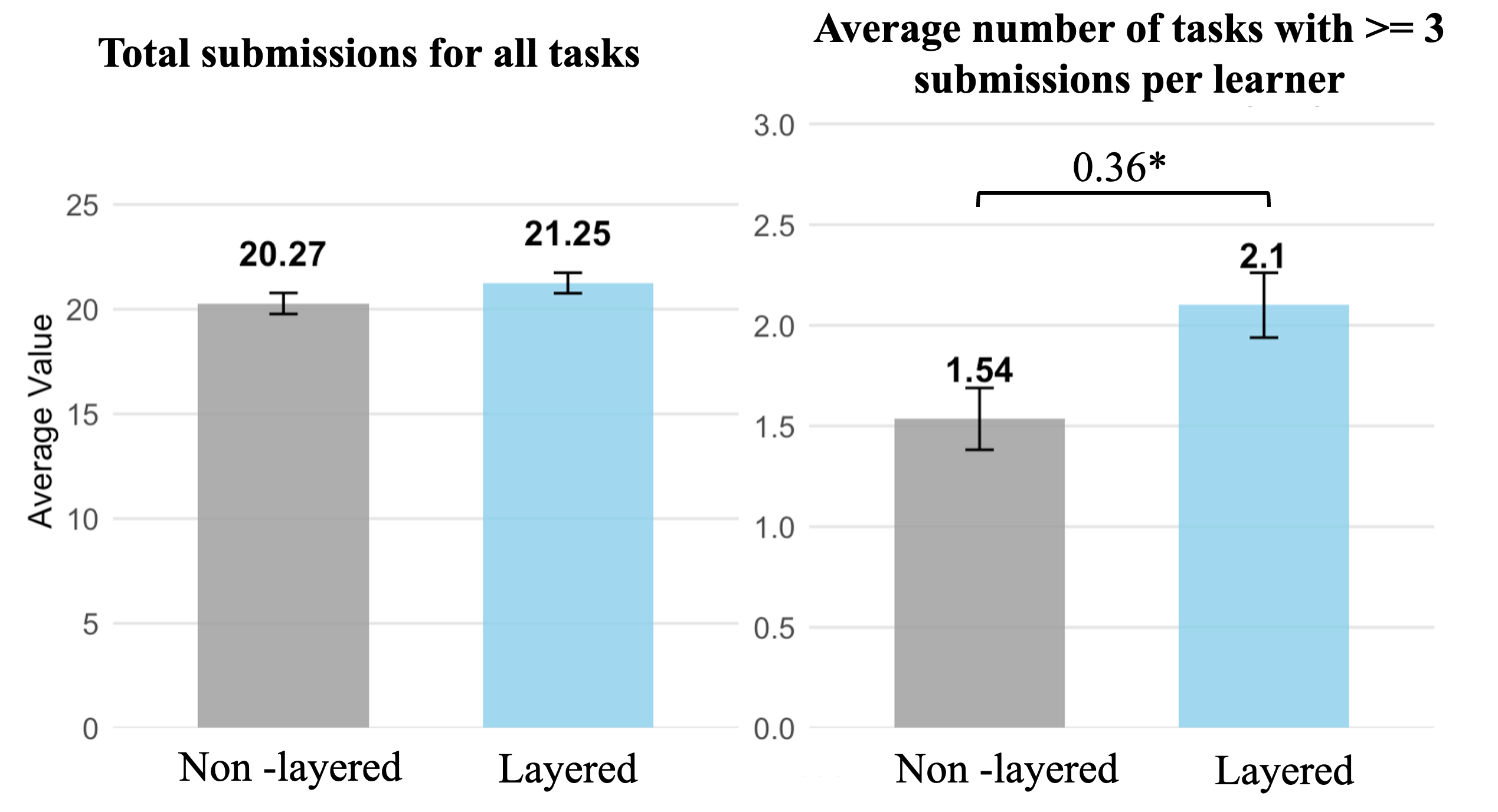}
  \caption{Comparison of total submissions and the number of tasks with multiple attempts between groups}
  \label{fig7}
\end{figure}

	\textbf{(b). Number of submissions and tasks}
Next, we examined the total number of responses submitted by learners (Figure \ref{fig7}, left). On average, learners in the non-layered feedback group submitted $20.3$ responses ($SD = 5.0$), while those in the layered feedback group submitted $21.3$ responses ($SD = 4.9$). An independent samples $t$-test showed no statistically significant difference between the two groups $t(197) = 1.4, p = .16, d = 0.2$.  To further capture learners’ persistence, we calculated the average number of tasks (out of $13$) for which each learner submitted more than two responses. This metric serves as an indicator of whether learners engaged in deeper exploration beyond initial submissions and basic corrections. As shown in right side of Figure \ref{fig7}, learners in the layered feedback group submitted three or more responses on significantly more tasks ($M = 2.1$) than those in the non-layered group ($M = 1.5$), $t(197) = 2.5, p = .012, d = 0.36$. These results suggest that layered feedback encouraged learners to revisit and refine their answers across a broader range of learning tasks.

	\textbf{(c). Modification}
Modification behavior serves as a critical indicator of how students internalize and act upon feedback. For open-ended questions, the non-layered group ($M = 17.8, SD = 15.0$) has higher edit distance than the layered group ($M = 15.0, SD = 11.0$), but it did not reach statistical significance ($U = 3857.5, p = .442$). 

% Collectively, these findings offer a nuanced perspective on learner behavioral engagement. Participants in the layered feedback group exhibited greater persistence: characterized by slightly longer engagement times, a higher frequency of submissions, and a more iterative revision process. In contrast, the non-layered feedback group demonstrated greater depth of content editing for open-ended questions, but this difference was not statistically significant.. 

 \subsection{Learners perceived layered feedback better, albeit with increased mental effort}

\begin{table}[t]
\centering
\footnotesize
\caption{Between-group comparison of learners' perceptions (Layered (N=100) vs. Non-layered (N=99))}
\label{tab:rq2-between}
\renewcommand{\arraystretch}{1.2}
\begin{tabular}{lcccccccc}
\toprule
 & \multicolumn{2}{c}{Mean} & & & & \multicolumn{2}{c}{95\% CI of $\Delta$} & \\
\cmidrule(lr){2-3} \cmidrule(lr){7-8}
\textbf{Variable} & Layered & Non & $t$ & $df$ & $p$ & High & Low & $d$ \\
\midrule
Perceived Specificity   & 4.12 & 4.28 & -1.66 & 184.9 & 0.1   & 0.03 & -0.35 & -0.23 \\
Perceived Encouragement & 4.39 & 3.95 & 3.86  & 147.7 & 0.000***& 0.68 & 0.22  & 0.55  \\
Perceived Independence  & 3.67 & 3.32 & 3.20  & 153.7 & 0.002**& 0.56 & 0.13  & 0.45  \\
Mental Effort           & 6.84 & 6.17 & 2.38  & 194.1 & 0.018*& 0.22 & 0.11  & 0.34  \\
\bottomrule
\multicolumn{9}{l}{* indicates $p < 0.05$; ** indicates $p < 0.01$; *** indicates $p < 0.001$.}
\end{tabular}
\end{table}

We compared the difference between the layered feedback and the non-layered feedback group among three perception dimensions and mental effort (as shown in Table \ref{tab:rq2-between}). Results showed that there was no significant difference between the two groups in perceived specificity, but significant differences emerged in other dimensions. Learners in the layered feedback group reported significantly higher perceived encouragement (p< .001) and higher perceived independence (p= .002), but also higher mental effort (p= .018). 

Within the layered feedback group, we also compared their perceptions of the first-layer feedback and the later-layer feedback they have received. Paired t-test results are shown in Table \ref{tab:rq2-within}. Again, there was no difference in perceived specificity, but learners reported significantly higher perceived encouragement (p< .001) and perceived independence (p< .001) in the first layer compared to the latter layer. 

Overall, these findings indicate that learners generally perceived layered feedback as providing more encouragement and supporting greater independence, albeit at the cost of higher reported mental effort.

%====== 2：Within-layered（paired t）======

\begin{table}[t]
\centering
\footnotesize
\caption{Within-layered comparison: first-layer vs. later-layer feedback (paired t) (N=100)}
\label{tab:rq2-within}
\setlength{\tabcolsep}{6pt}
\renewcommand{\arraystretch}{1.2}
\begin{tabular}{lcccccccc}
\toprule
 & \multicolumn{2}{c}{Mean} & & & & \multicolumn{2}{c}{95\% CI of $\Delta$} & \\
\cmidrule(lr){2-3} \cmidrule(lr){7-8}
\textbf{Variable} & First & Later & $t$ & $df$ & $p$ & Low & High & $d$ \\
\midrule
Perceived Specificity   & 4.05 & 4.19 & -1.56 & 99 & 0.12 & -0.31 & 0.04 & -0.16 \\
Perceived Encouragement & 4.54 & 4.25 & 3.48  & 99 & 0.000*** & 0.12 & 0.46 & 0.35 \\
Perceived Independence  & 4.04 & 3.30 & 6.69  & 99 & 0.000*** & 0.52 & 0.96 & 0.67 \\
\bottomrule
\multicolumn{9}{l}{* indicates $p < 0.05$; ** indicates $p < 0.01$; *** indicates $p < 0.001$.}
\end{tabular}
\end{table}

\subsection{Mediation Analysis of Feedback Mechanisms}
\begin{figure}[h]
  \centering
  \includegraphics[trim={0cm 0cm 0 0cm}, clip,width=0.9\linewidth]{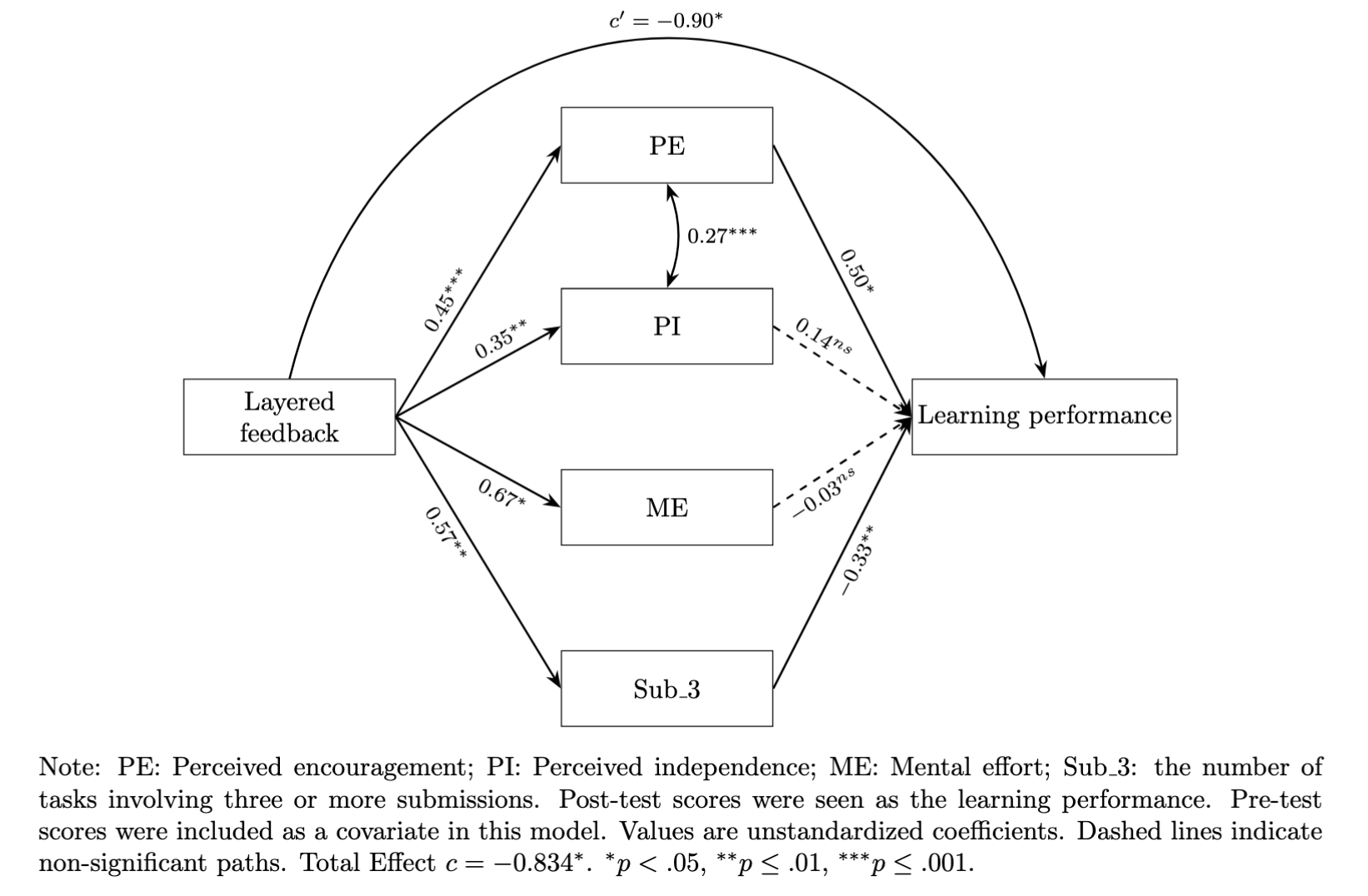}
  \caption{Mediation model of how layered feedback influence learning performance compared to non-layered feedback}
  \label{fig8}
\end{figure}

To further uncover the underlying mechanisms through which the feedback version influenced learning performance, a multiple mediation analysis was performed using Perceived Encouragement (PE), Perceived Independence (PI), Mental effort (ME), and average number of tasks with $\geq 3$ submissions (sub\_3) as mediators. Those four mediators were selected because they were significantly different across the two feedback conditions (see RQ2 and RQ3 results),  representing learners' affective (PE and PI), cognitive (ME) perceptions, and behavioral engagement (sub\_3), respectively. The results of the path analysis are illustrated in Figure \ref{fig8} and Table \ref{tab:mediation_results_full}.

The overall goodness-of-fit of the mediation model was evaluated using multiple fit indices. The chi-square test was significant ($\chi^2(9) = 17.456, p = 0.042$). Other practical fit indices indicated an acceptable to good fit for the data \citep{bentler1990comparative}. Specifically, the Comparative Fit Index (CFI) was 0.952 (exceeding the 0.95 threshold for excellent fit), and the Standardized Root Mean Square Residual (SRMR) was 0.061 (below the 0.08 threshold). The Root Mean Square Error of Approximation (RMSEA) was 0.069, close to 0.06. While the Tucker-Lewis Index (TLI = 0.893) was marginally below the 0.90 cutoff, the holistic evaluation of all indices suggests that the hypothesized model adequately represents the empirical data.

\subsubsection{Direct Effects of the Feedback Conditions on Mediators and Covariances of Mediators}

As seen in Figure \ref{fig8}, all the direct connections between feedback version and mediators were statistically significant. This result is not surprising given the mediators were selected based on significant differences by conditions. 
In addition, since PE and PI are self-reported affective-related variables that are correlated (p<.001), we also added bidirectional link between them in our mediation model. .

\subsubsection{Predictors of Learning Performance}
When controlling for the treatment and pre-test scores, the four mediators showed divergent relationships with the learning performance (Post-test scores):
\begin{itemize}
\item Perceived Encouragement positively predicted learning outcomes ($B=0.50,SE=0.24,p=.04$), suggesting that students who felt more encouraged tended to achieve higher scores.
    \item Average number of tasks with $\geq 3$  submissions negatively predicted learning outcomes ($B=-0.33,SE=0.10, p=.001$). Higher submission counts were associated with lower post-test scores, potentially indicating trial-and-error behavior rather than constructive iteration.
    \item Mental Effort ($B=-0.03,SE=0.09, p=.77$) and Perceived Independence ($B=0.14,SE=0.26, p=.59$) did not significantly predict post-learning performance.
\end{itemize}

\subsubsection{Indirect Effects (Mediation Analysis)}
The analysis revealed two significant but opposing indirect paths, indicating a competitive mediation mechanism (see Table~\ref{tab:mediation_results_full}):
\begin{itemize}
    \item The Positive and Neutral Affective Perception Paths: There was a significant positive indirect effect through perceived encouragement ($Ind = 0.22$, $SE = 0.12$, $p = .06, 95\%CI[0.014,0.467]$). This result suggests that the layered feedback improved performance by fostering a sense of encouragement and validation in learners. However, there is no significant positive indirect effect through perceived independence ($Ind = 0.05$, $SE = 0.10$, $p = .61, 95\%CI[-0.126,0.253]$). 
    \item The Negative Behavioral Path: Conversely, there was a significant negative indirect effect through the average number of tasks with >=3 submissions  ($Ind = -0.19$, $SE = 0.09$, $p = .03, 95\%CI[-0.374,-0.037]$). This result suggests that while the intervention prompted more actions (submissions), this increased frequency was detrimental to the final score, likely reflecting inefficient learning strategies.
    \item The Neutral Cognitive Path: The indirect effect through mental effort was not significant ($Ind = -0.02$, $SE = 0.06$, $p = .80,95\%CI[-0.167,0.099]$), indicating that self-reported mental effort did not translate into a performance advantage or disadvantage in this context.
\end{itemize}

\subsubsection{Total Effects}
The total indirect effect was not significant ($TotalInd=0.07, SE=.16, p=0.67, 95\%CI[-0.248, 0.387]$) because the opposing pathways cancel each other out. However, there was a statistically significant negative direct effect, reflecting another cause not captured by the mediators. As a result, the total effect of the layered feedback on learning outcomes was statistically significant and negative ($B = -0.83$, $SE = 0.36$, $z = -2.30$, $p = .02, [-1.545, -0.117]$). This result indicates that despite the positive emotional support, the experimental group scored significantly lower overall, driven by the negative behavioral patterns and other direct effects.

\begin{table}[htbp]
\centering
\small
\caption{Path Coefficients and Indirect Effects of the Mediation Model}
\label{tab:mediation_results_full}
\begin{tabular}{p{6cm} ccccc}
\toprule
\textbf{Path / Effect} & \textbf{Coeff. ($B$)} & \textbf{$SE$} & \textbf{$z$-value} & \textbf{$p$-value} & \textbf{95\% CI} \\
\midrule
\textit{Component Paths (Regression Results)} & & & & & \\
Layered feedback $\rightarrow$ PE ($a_1$)& 0.45& 0.12& 3.86 & .000& $[0.225, 0.678]$ \\
Layered feedback  $\rightarrow$ PI ($a_2$) & 0.35& 0.11& 3.24 & .001 & $[0.137, 0.566]$ \\
Layered feedback  $\rightarrow$ Mental Effort ($a_3$) & 0.67& 0.28& 2.36 & .02& $[0.119, 1.229]$ \\
Layered feedback  $\rightarrow$ Sub\_3 ($a_4$) & 0.57& 0.22& 2.57 & .01& $[0.134, 0.995]$ \\
PE $\rightarrow$ Post\_test ($b_1$) & 0.50& 0.24& 2.06 & .04& $[0.034, 0.994]$ \\
PI $\rightarrow$ Post\_test ($b_2$) & 0.14& 0.26& 0.54 & .59& $[-0.356, 0.642]$ \\
Mental Effort $\rightarrow$ Post\_test ($b_3$) & -0.03& 0.09& -0.29 & .77& $[-0.194, 0.137]$ \\
Sub\_3 $\rightarrow$ Post\_test ($b_4$) & -0.33& 0.10& -3.40 & .001 & $[-0.521, -0.140]$ \\
Direct Effect: Layered feedback  $\rightarrow$ Post\_test ($c'$) & -0.90& 0.38& -2.40 & .02& $[-1.618, -0.158]$ \\
\midrule
\textit{Covariances} & & & & & \\
PE $\leftrightarrow$ PI & 0.27& 0.07& 4.16 & .000 & $[0.145, 0.401]$ \\
\midrule
\textit{Mediation Effects (Bootstrap Results)} & & & & & \\
Indirect via PE ($ind_{PE}$) & \textbf{0.22}& 0.12& 1.89 & .06& \textbf{$[0.014, 0.467]*$}\\
Indirect via PI ($ind_{PI}$) & 0.05& 0.10& 0.51 & .61& $[-0.126, 0.253]$ \\
Indirect via Mental Effort ($ind_{ME}$) & -0.02& 0.06& -0.26 & .80& $[-0.167, 0.099]$ \\
Indirect via Sub\_3 ($ind_{Sub}$) & \textbf{-0.19}& 0.09& -2.12 & \textbf{.03*}& \textbf{$[-0.374, -0.037]$} \\
Total Indirect Effect & 0.07& 0.16& 0.43 & .67& $[-0.248, 0.387]$ \\
\textbf{Total Effect ($c$)} & \textbf{-0.83}& \textbf{0.36}& \textbf{-2.30} & \textbf{.02*}& \textbf{$[-1.545, -0.117]$} \\
\bottomrule
\multicolumn{6}{p{15.5cm}}{\small \textit{Note.} $N = 199$. Coeff. = Unstandardized coefficients. CI = Bias-corrected bootstrap confidence intervals (based on 5000 samples). Pre-test scores were included as covariates. Significant specific indirect and total effects are highlighted in bold. * $p < .05$. Note that the significance of indirect effects is robustly determined by the 95\% CI excluding zero, confirming the mediation of PE despite its $p$-value.} \\
\end{tabular}
\end{table}

% discussion and conclusion (2 to 3 pages)
\section{Discussion}
 \subsection{The non-layered feedback condition yielded significantly greater learning performance than the layered feedback condition}
First, we found that the non-layered feedback condition yielded significantly greater post-learning performance. This result aligns with prior evidence \citep{roper1977feedback,swart2019supporting,butler2013explanation,wang2024make,van2015effects}, which consistently demonstrates that feedback containing correct answers accompanied by explanations or rationales is recommended for optimal student learning. However, these prior studies primarily contrasted elaborated feedback with simple correctness judgments or correct-answer-only conditions. By introducing a new comparison between layered feedback (learner-centered feedback followed by elaborated feedback with the correct answer) and non-layered feedback (only elaborated feedback with the correct answer), the present study contributes new empirical evidence that again confirms the effectiveness of elaborated feedback with the correct answer. Nevertheless, this result contrasts with theoretical perspectives emphasizing learner-centered feedback \citep{carless2022teacher,spooner2022self,Ryan18082021,butler2008effects}. According to these perspectives, feedback that fosters affect, independence, and self-regulated learning should enhance outcomes by granting students greater agency. Rather than passively receiving feedback, learners are expected to engage actively with it, reflect on their errors, and construct new understanding. One possible explanation concerns the nature of the learning outcome measure used in this study, which primarily assessed learners’ understanding and application of the multimedia principle. In such contexts, elaborated feedback with correct answers and explanations may be particularly effective in supporting conceptual mastery. However, in more advanced learning scenarios that target higher-order thinking skills, feedback emphasizing emotional support, agency, independence, and self-regulation may prove more beneficial \citep{scherer2024effective}.  In addition, the divergence between theory and our empirical findings may also be explained by results related to learners' affective, behavioral, and cognitive engagement.   

\subsection{The Dual Nature of the Layered Feedback}
Our results revealed a complex, ``double-edged'' impact of layered feedback on learning outcomes, characterized by two mediation pathways with opposing effects. 

\subsubsection{The Positive Affective Path: Encouragement as a Motivator}
Consistent with prior research on supporting learner motivation (e.g., \citep{Dai02012025}), the ``Positive Affective Path'' demonstrates that our feedback design successfully conveyed a sense of support, which also aligns with the findings of \citep{alsaiari2025emotionally}. The significant positive indirect effect via perceived encouragement suggests that when students feel validated and supported by the feedback system, their learning outcomes improve. Feedback tone can be important, particularly in creating a psychological safety net that is conducive to learning. However, the other possible motivational pathway through perceived independence was not significant, potentially because perceived independence is significantly correlated with perceived encouragement.

\subsubsection{The Negative Behavioral Path}
However, the benefits of the affective path were neutralized by the ``Negative Behavioral Path.'' The layered feedback led to a significantly more tasks with three or more submissions, which negatively predicted post-test scores, aligning with some studies in the context of ITS that a high frequency of submissions may indicate ``gaming the system'' rather than constructive iteration \citep{Baker2004}. Similar studies have found that students tend to rapidly click through or pay minimal attention to initial hints in order to reach the bottom-out hints directly \citep{mathews2008analysing,aleven2000limitations}, a behavior that leads to less effective learning.  Given that total learning time did not differ significantly between the two feedback conditions, the increased submission count implies that students dedicated less time to processing each feedback instance. The prevalence of tasks with three or more submissions indicates that students often needed the additional feedback from the second layer. However, students did not benefit the first attempt, perhaps because they moved too quickly through rapid trial-and-error behavior. That is, they may have engaged in shallow cognitive processing, where students prioritize task completion over see feedback as the scaffold to learn \citep{aleven2016help,Vincent2003}.

\subsubsection{The Neutral Cognitive Path: A Paradox of Ineffective Mental Effort}
Another interesting finding involves the ``Neutral Cognitive Path.'' The layered feedback significantly increased students' mental effort, yet this effort did not translate into better learning outcomes, creating a paradox of ineffective effort. Students perceived that they were actively engaged (high effort), but this engagement may have been misdirected toward navigating the interface or processing overwhelming information in the feedback, rather than deeply understanding or taking up feedback. This suggests that high mental effort is not always a proxy for high germane cognitive load (the load devoted to learning); in this case, it might have been extraneous load or effort spent on inefficient strategies \citep{schnotz2007reconsideration}. Furthermore, the mental effort analyses relied on self-reports, which might not accurately represent students' true cognitive engagement. To more precisely capture cognitive, future studies should incorporate more objective measures such as cued-retrospective or think aloud protocols.

Final, layered feedback exhibited a substantial negative direct effect on post-test performance that remained unaccounted for by the current mediators. This highlights a critical gap in our understanding of how layered feedback influences learning outcomes. Future studies should move beyond self-reported perceptions to collect more nuanced, time-series data, such as eye-tracking or think-aloud data when receiving feedback, to reveal the hidden dynamics that undermine performance.

\section{Implications for LLM-generated Feedback Design}

These findings offer critical insights for the design of future LLM-generated feedback systems for online higher education. The goal is to decouple the positive affective effects from the negative behavioral consequences. Future iterations should aim to:

\textbf{a. Retain encouraging tones: }The value of perceived encouragement is clear and should be preserved. Our results confirm that the ``encouraging'' tone of LLM-generated feedback significantly enhances perceived encouragement, which then enhanced academic performance. Future LLM-generated feedback designs should preserve and refine this affective scaffolding, maintaining an empathetic, encouraging persona (e.g., ``Don't worry, this is a tricky concept...'', ``you are on the right track, I see you'') to reduce anxiety and maintain motivation, particularly for struggling learners, as positive affect is a precursor to persistence. However, an encouraging tone (polite tone) may increase feedback length, and this additional information may overwhelm the receiver \citep{jug2019giving}. Combined with this insight, feedback should also be concise, to maintain a balance between supportive tone and concise information. 

\textbf{b. Constrain behavioral gaming:}  To mitigate the negative effects of gaming the system without meaningfully engaging with first-layer feedback, researchers and designers may consider several design strategies. One approach is to regulate the timing of resubmissions. For example, imposing a mandatory delay before a second attempt could prompt learners to attend to and process the first-layer feedback, rather than treating it as an obstacle to bypass. Another alternative approach is to allow a certain degree of gaming within the inner loop (i.e., iterative attempts within a task), while strategically structuring the outer loop (i.e., task sequencing) \citep{vanlehn2006behavior,aleven2015beginning}. In this design, even if students engage in superficial optimization within individual tasks, the sequenced arrangement of tasks can reinforce key concepts over time. By aligning tasks in a coherent progression, designers can leverage repeated engagement across contexts to promote cumulative learning, thereby mitigating the potential negative consequences of gaming behaviors.

\textbf{c. Reduce overwhelming load and facilitate constructive cognitive processing}: 
One possible explanation for the limited effectiveness of layered feedback is that it may provide excessive information, thereby overwhelming learners and constraining meaningful engagement, even when reflective questions are included. In other words, the cognitive load imposed by dense feedback may inhibit learners from processing the information deeply and productively. To facilitate more constructive cognitive processing, future designs should shift from a model of information transmission to one of constructive interaction \citep{Ryan18082021}. Rather than unidirectionally delivering feedback for students to read, designers could adopt a ``feedback-and-respond'' approach. In this model, learners would first respond to the reflective prompts embedded in the feedback or generate self-explanations based on the feedback content. Such bidirectional interaction would encourage active processing and deeper engagement before resubmission. This structured interaction may better support meaningful revision and learning outcomes.

\section{Conclusion}
This study compared LLM-generated layered and non-layered feedback to examine their effects on learning performance and behind mechanisms. Findings revealed that while layered feedback led to higher perceived encouragement, perceived independence, mental effort as well as behavioral engagement, there is a competitive mediation mechanism. Non-layered feedback yielded greater performance gains. Overall, the results highlight a trade-off between learning gains, behavioral engagement and greater perceptions. 

This study has three limitations. First, the learning-by-doing tasks were relatively low-level and may not have fully engaged higher-order thinking; future research should test feedback designs in tasks requiring more advanced cognition and use a delayed post-test to test the long-term learning. Second, the study focused on a single course context, and effects may differ across disciplines. Third, the process data relied mainly on log traces; future work could integrate richer measures (e.g., eye-tracking) to better capture how learners attend to and process feedback, and get a nuanced analysis of the learning process.

Despite some limitations, this study yields findings of theoretical, methodological and practical significance. At the theoretical level, the results challenge the assumption that layered feedback is superior to non-layered feedback. The data demonstrate that layered feedback did not outperform direct feedback in terms of learning gain. This suggests that the impact of feedback is not uniformly positive and may be moderated by learner feature, task characteristics, and the time frame of measurement, thereby enriching our understanding of feedback mechanisms. Methodologically, this study went beyond performance outcomes by incorporating questionnaire responses, log behaviors, and cognitive load measures. This multidimensional approach reveals the tension between learning outcomes, processes, and experiences, contributing to a shift in feedback research from a purely outcome-oriented perspective toward a more holistic view. Practically, the findings provide concrete implications for teachers and the design of LLM-based educational feedback: direct feedback appears more efficient when the instructional goal is short-term performance in online higher education, whereas layered feedback retains unique value in fostering motivation, independence, and long-term development.



\bibliographystyle{elsarticle-harv}
\bibliography{Ref}
\appendix


\section*{Supplementary material}

# Prompt_Appendix

## 1. Prompt for elaborated feedback with correct answer

### 1.1 Version for multiple-choice questions

You are tasked with generating clear, effective feedback for a student's multiple-choice answer and then formatting it into a structured JSON output. Complete both tasks in sequence.

### Task 1: Generate Feedback

Generate feedback that meets all five criteria:

**Required Criteria:**

1. **Judgment Statement**: Begin by clearly stating whether the student's answer is correct or incorrect.

2. **Explain the Student's Answer with Context**:

   - If incorrect: Provide the correct answer directly to the student, and explain why this answer is correct and why the one they chose is incorrect.

   - If correct: Briefly explain why their choice fits and reference specific elements from the question.

3. **Use Specific Details from the Question**: Connect explanations to concrete elements from the question scenario—avoid abstract or generalized definitions.

4. **Provide suggestions for further study**: Include at least one strategy to help the student improve on similar future questions. This can be:

   - A practical tip (e.g., "Try asking whether this shows appearance or relationships.")

   - A strategic suggestion for solving similar problems (e.g., "Take a look at...")

- A suggestion to help students remember key knowledge or review specific materials (e.g., "Review the concept of...")

- A reflective question (e.g., "What kind of information does this help you understand?")

5. **Clarity and Brevity**: Keep the entire feedback clear, constructive, and under 100 words.

**Question Details:**

- Multiple choice question: {According to the multimedia principle, learning is better when a lesson:}

- Choices: {A. Is delivered on a computer; B. Adds graphics to a text description; C. Adds text to a graphic illustration; D. Uses animation rather than still visuals}

- Retrieved related slides: {as shown in the uploaded version}

- Student's Response: {D}

- Correct Answer: {B}

# Task 2: Format Output

After generating the feedback, format your response as a JSON object with this exact structure:

```
{
  "score": "[0 for incorrect, 1 for correct]",
  "feedback": "[A clear, concise revision of the original feedback, retaining key points and removing redundancy. Tooltips are integrated as plain terms.]",
  "terms": [
    {
      "[term 1]": "[tooltip explanation]"
    },
    {
      "[term 2]": "[tooltip explanation]"
    }
  ],
```

```
  "quotes": [
    {
      "section": "statement",
      "quotes": [
        "[Direct, concise phrases stating correctness or incorrectness.]"
      ]
    },
    {
      "section": "explanation",
      "quotes": [
        "[Phrases explaining the mistake or correct reasoning.]"
      ]
    },
    {
      "section": "advice",
      "quotes": [
        "[Concrete improvement suggestions or reflective prompts.]"
      ]
    }
  ]
}
```

**Formatting Instructions:**

- First, Identify terms from the feedback that require explanation (key concepts or technical terms) and extract their tooltip-style explanations

- Do not repeat tooltip details within the feedback body

- Extract strictly quotable phrases from the concised feedback, categorized into:

  ○ **statement**: Phrases about whether the answer is correct or incorrect

  ○ **explanation**: Reasoning that explains the mistake or correct logic

  ○ **advice**: Actionable suggestions for improvement
    CRITICAL: The terms in the "terms" array must use the exact wording as it appears in the feedback text, with no modifications, paraphrasing, or

rewording whatsoever
**Final Output**: Provide only the JSON object in the exact format specified above. No additional explanation, comments, or plain text are allowed.

# 1.2 Version for open-ended questions

You are tasked with generating clear, effective feedback for a student's open-ended answer and then formatting it into a structured JSON output. Complete both tasks in sequence.

## Task 1: Generate Feedback

Generate feedback that meets all five criteria:

**Required Criteria:**

1. **Judgment Statement**: Begin by clearly stating whether the student's answer is correct, incorrect, or partially correct.
   Note: Please evaluate the student's response to the open-ended question by comparing it with the reference answer:
   • Score 1 (Correct): The response fully covers all key points in the reference answer, with no contradictions.
   • Score 2 (Partially Correct): The response includes some of the key points but misses others or contains information that partly contradicts the reference.
   • Score 0 (Incorrect): The response does not cover any key points from the reference answer or directly contradicts them.

2. **Explain the Student's Answer with Context**:

   • If incorrect: Provide the correct answer directly to the student, and explain why this answer is correct and why the one they provided is incorrect.

   • If correct: Briefly explain why their answer is accurate and reference specific elements from the question.

   • If partially correct: Acknowledge what aspects are correct, then explain what is missing or incorrect, and provide the complete correct answer.

3. **Use Specific Details from the Question**: Connect explanations to concrete elements from the question scenario—avoid abstract or generalized

definitions.

4. **Provide suggestions for further study**: Include at least one strategy to help the student improve on similar future questions. This can be:

   - A practical tip (e.g., "Try asking whether this shows appearance or relationships.")

   - A strategic suggestion for solving similar problems (e.g., "Take a look at...")

   - A suggestion to help students remember key knowledge or review specific materials (e.g., "Review the concept of...")

   - A reflective question (e.g., "What kind of information does this help you understand?")

5. **Clarity and Brevity**: Keep the entire feedback clear, constructive, and under 100 words.

**Question Details:**

- Open-ended question: {Recognize when the multimedia principle has been violated and when it has been applied well. Identify whether there are any part(s) of the lesson that represent a good application and/or violation of the multimedia principle. Provide specific reasons to support your judgment. A teacher is explaining the causes of World War I. On the slide, there's a map of Europe in 1914 and a detailed timeline of alliances forming. At the same time, an image of a famous WWI painting (a battlefield with soldiers and explosions) slowly fades in for visual impact. The teacher is narrating the complex political tensions and alliances.}

- Retrieved related slides: {as shown in the uploaded version}

- Student's Response: {I think this is a good application because it combines text and images.}

- Model Answer/Key Points: {Good Application: The map of Europe and timeline effectively support learning by providing relevant visual information that helps students understand the complex political relationships and sequence of events leading to WWI. Violation: The battlefield painting represents extraneous visual content that doesn't support the learning objective about

causes of the war. This decorative image creates unnecessary cognitive load and distracts from the core instructional content about political tensions and alliances.}

## Task 2: Format Output

After generating the feedback, format your response as a JSON object with this exact structure:

```
{
  "score": "[0 for incorrect, 1 for correct, 2 for partially correct]",
  "feedback": "[A clear, concise revision of the original feedback, retaining key points and removing redundancy. Tooltips are integrated as plain terms.]",
  "terms": [
    {
      "[term 1]": "[tooltip explanation]"
    },
    {
      "[term 2]": "[tooltip explanation]"
    }
  ],
  "quotes": [
    {
      "section": "statement",
      "quotes": [
        "[Direct, concise phrases stating correctness, incorrectness, or partial correctness.]"
      ]
    },
    {
      "section": "explanation",
      "quotes": [
        "[Phrases explaining the mistake, correct reasoning, or missing elements.]"
      ]
    },
```

```
  {
    "section": "advice",
    "quotes": [
      "[Concrete improvement suggestions or reflective prompts.]"
    ]
  }
 ]
}
```

**Formatting Instructions:**

- First, identify terms from the feedback that require explanation (key concepts or technical terms) and extract their tooltip-style explanations

- Do not repeat tooltip details within the feedback body

- Extract strictly quotable phrases from the concised feedback, categorized into:

  - **statement**: Phrases about whether the answer is correct, incorrect, or partially correct

  - **explanation**: Reasoning that explains the mistake, correct logic, or missing elements

  - **advice**: Actionable suggestions for improvement
    CRITICAL: The terms in the "terms" array must use the exact wording as it appears in the feedback text, with no modifications, paraphrasing, or rewording whatsoever
    **Final Output**: Provide only the JSON object in the exact format specified above. No additional explanation, comments, or plain text are allowed.

# 2. Prompt for Learner-centered feedback

## 2.1  Version for multiple-choice questions

You are tasked with generating clear, effective feedback for a student's multiple-choice answer and then formatting it into a structured JSON output. Complete both tasks in sequence.

## Task 1: Generate Feedback

Generate feedback that meets all seven criteria:

**Required Criteria:**

1. **Judgment Statement**: Begin by clearly stating whether the student's answer is correct or incorrect.

2. **Explain the Student's Answer with Context**:

   - If incorrect: **Start by directly quoting or paraphrasing the student's specific response**, then explain what their answer means and why it doesn't fully address the question requirements. **CRITICAL: DO NOT reveal, mention, hint at, or describe the correct answer in any form.**

   - If correct: **Reference their specific response**, briefly explain why their choice fits and reference specific elements from the question.

3. **Use Specific Details from the Question**: Connect explanations to concrete elements from the question scenario—avoid abstract or generalized definitions. **MANDATORY: Reference at least 3 specific elements mentioned in the question prompt** (e.g., "the map of Europe in 1914," "detailed timeline of alliances," "WWI battlefield painting," "teacher narrating political tensions"). **Connect the student's response directly to these specific scenario elements.**

4. **Provide Multiple Suggestions for Further Study**: Include several strategies to help the student improve on similar future questions. **MUST incorporate specific references to the retrieved slides content and question context. CRITICAL: Must include at least 2-3 specific reflective questions that guide the student's thinking process**:

   - **Context-specific reflective questions** that guide critical thinking about the specific scenario presented. **MANDATORY: Generate 2-3 very specific reflective questions directly related to this question's scenario and content** (e.g., "When you see the Microsoft Outlook screen capture

showing menus and toolbars, ask yourself: What is this image helping you understand—the structure, the appearance, or the relationships?" "Does this screen capture organize information or show what something looks like?")

- **Slide-informed learning strategies** that reference relevant concepts from the retrieved slides (e.g., connect to principles, frameworks, or categorizations shown in the slide content)

- **Scenario-based exploration prompts** that encourage students to examine similar examples or apply the same analytical framework to comparable situations

- **Question-specific metacognitive prompts** that help students develop critical thinking skills for evaluating the specific type of content or scenario presented

5. **Encourage Student Autonomy and Agency**:

   - Invite students to take ownership of their learning

   - Suggest they seek additional resources or engage in independent study

   - Encourage them to ask questions and engage in dialogue

   - Frame learning as an active, student-driven process

   - **Connect to specific slide content and encourage independent exploration of the multimedia principle**

6. **Maintain Highly Positive and Supportive Tone**: Use encouraging language that:

   - Celebrates their thinking process and effort

   - Frames mistakes as learning opportunities

   - Builds confidence and motivation

   - Recognizes their potential for growth

   - Values their engagement and curiosity

7. **Clarity and Engagement**: Keep the entire feedback clear, constructive, and under 120 words in a single paragraph format that feels conversational and

supportive. **Must include at least 2-3 specific references to elements from the question scenario and 1-2 references to the slide content. Include 2-4 key technical terms that would benefit from tooltip explanations. CRITICAL: Must include at least 2 specific reflective questions that guide student thinking about this particular scenario.**

**ABSOLUTE RESTRICTION**:

- **NEVER provide, hint at, or describe the correct answer**

- **NEVER mention specific elements that should or shouldn't be included**

- Focus ONLY on guiding the student's thinking process and self-discovery

**REQUIRED INTEGRATION**:

- **MUST reference specific elements from the question scenario** (using the concrete details, examples, or contexts provided in the question prompt)

- **MUST incorporate guidance from the retrieved slides** (referencing relevant principles, concepts, frameworks, or information shown in the slide content)

- **MUST make suggestions contextually relevant** to the specific scenario and learning context rather than generic advice

**Question Details:**

- Multiple choice question: {A lesson on Microsoft Outlook includes a screen capture to illustrate the menus and tool bars. The screen capture is an example of a(n) _____ graphic}

- Choices: {Decorative; Representational; Organizational; Relational}

- Retrieved related slides: {as shown in the uploaded version}

- Student's Response: {Organizational}

- Correct Answer: {Representational}

## Task 2: Format Output

After generating the feedback, format your response as a JSON object with this exact structure:

```
{
  "score": "[0 for incorrect, 1 for correct]",
  "feedback": "[A clear, concise revision of the original feedback, retaining key
points and removing redundancy. Tooltips are integrated as plain terms. NEVE
R reveal the correct answer or model answer elements.]",
  "terms": [
    {
      "[term 1]": "[tooltip explanation]"
    },
    {
      "[term 2]": "[tooltip explanation]"
    }
  ],
  "quotes": [
    {
      "section": "statement",
      "quotes": [
        "[Direct, concise phrases stating correctness or incorrectness.]"
      ]
    },
    {
      "section": "explanation",
      "quotes": [
        "[Phrases explaining why the student's response doesn't fully address th
e question, WITHOUT revealing the correct answer or model answer element
s.]"
      ]
    },
    {
      "section": "advice",
      "quotes": [
        "[Concrete improvement suggestions or reflective prompts.]"
      ]
    }
```

```
    ]
}
```

**Formatting Instructions:**

- **CRITICAL**: Maintain the same ABSOLUTE RESTRICTION as Task 1 - DO NOT reveal, mention, hint at, or describe the correct answer, model answer, or any specific elements from the model answer during formatting

- **Terms Extraction**: Please identify 1-3 key technical terms or concepts from the feedback that students might need clarification on (e.g., "multimedia principle", "cognitive load", "relevance", "learning objectives", "decorative", etc.). The terms must use the exact wording as it appears in the feedback text.

- **Quote Extraction Guidelines**:

  - **statement**: Extract the direct judgment phrase(s) about correctness (e.g., "Your answer is incorrect", "partially correct", "correct")

  - **explanation**: Extract ALL sentences that explain the student's response and provide reasoning about why it doesn't fully address the question. Include complete sentences and connected ideas, not just fragments. This should capture the full reasoning without revealing correct answers.

  - **advice**: Extract sentences that provide concrete suggestions, reflective questions, or learning strategies

- **Complete Sentence Extraction**: When extracting quotes, prioritize complete sentences or meaningful phrase clusters rather than sentence fragments. If multiple sentences form a cohesive explanation, include them together in a single quote.

**Final Output:** Provide only the JSON object in the exact format specified above. No additional explanation, comments, or plain text are allowed.

## 2.2 Version for open-ended questions

You are tasked with generating clear, effective feedback for a student's open-ended answer and then formatting it into a structured JSON output. Complete both tasks in sequence.

## Task 1: Generate Feedback

Generate feedback that meets all seven criteria:

**Required Criteria:**

1. **Judgment Statement**: Begin by clearly stating whether the student's answer is correct or incorrect.
   Note: Please evaluate the student's response to the open-ended question by comparing it with the reference answer:
   • Score 1 (Correct): The response fully covers all key points in the reference answer, with no contradictions.
   • Score 2 (Partially Correct): The response includes some of the key points but misses others or contains information that partly contradicts the reference.
   • Score 0 (Incorrect): The response does not cover any key points from the reference answer or directly contradicts them.

2. **Explain the Student's Answer with Context**:
   • If incorrect: **Start by directly quoting or paraphrasing the student's specific response**, then explain what their answer means and why it doesn't fully address the question requirements. **CRITICAL: DO NOT reveal, mention, hint at, or describe the correct answer in any form.**

   • If correct: **Reference their specific response**, briefly explain why their choice fits and reference specific elements from the question.

   • If partially correct: **Acknowledge their specific response and what aspects are correct**, then tell students that something is missing or incomplete in their analysis, but **CRITICAL: DO NOT reveal, mention, hint at, or describe the correct answer in any form.**

3. **Use Specific Details from the Question**: Connect explanations to concrete elements from the question scenario—avoid abstract or generalized definitions. **MANDATORY: Reference at least 3 specific elements mentioned in the question prompt** (e.g., "the map of Europe in 1914," "detailed timeline of alliances," "WWI battlefield painting," "teacher narrating political tensions"). **Connect the student's response directly to these specific scenario elements.**

4. **Provide Multiple Suggestions for Further Study**: Include several strategies to help the student improve on similar future questions. **MUST incorporate specific references to the retrieved slides content and question context. CRITICAL: Must include at least 2-3 specific reflective questions that guide the student's thinking process**:

   - **Context-specific reflective questions** that guide critical thinking about the specific scenario presented. **MANDATORY: Generate 2-3 very specific reflective questions directly related to this question's scenario and content** (e.g., "When you see the Microsoft Outlook screen capture showing menus and toolbars, ask yourself: What is this image helping you understand—the structure, the appearance, or the relationships?" "Does this screen capture organize information or show what something looks like?")

   - **Slide-informed learning strategies** that reference relevant concepts from the retrieved slides (e.g., connect to principles, frameworks, or categorizations shown in the slide content)

   - **Scenario-based exploration prompts** that encourage students to examine similar examples or apply the same analytical framework to comparable situations

   - **Question-specific metacognitive prompts** that help students develop critical thinking skills for evaluating the specific type of content or scenario presented

5. **Encourage Student Autonomy and Agency**:

   - Invite students to take ownership of their learning

   - Suggest they seek additional resources or engage in independent study

   - Encourage them to ask questions and engage in dialogue

   - Frame learning as an active, student-driven process

   - **Connect to specific slide content and encourage independent exploration of the multimedia principle**

6. **Maintain Highly Positive and Supportive Tone**: Use encouraging language that:

- Celebrates their thinking process and effort

- Frames mistakes as learning opportunities

- Builds confidence and motivation

- Recognizes their potential for growth

- Values their engagement and curiosity

7. **Clarity and Engagement**: Keep the entire feedback clear, constructive, and under 120 words in a single paragraph format that feels conversational and supportive. **Must include at least 2-3 specific references to elements from the question scenario and 1-2 references to the slide content. Include 2-4 key technical terms that would benefit from tooltip explanations. CRITICAL: Must include at least 2 specific reflective questions that guide student thinking about this particular scenario.**

**ABSOLUTE RESTRICTION**:

- **NEVER provide, hint at, or describe the correct answer**

- **NEVER mention specific elements that should or shouldn't be included**

- Focus ONLY on guiding the student's thinking process and self-discovery

**REQUIRED INTEGRATION**:

- **MUST reference specific elements from the question scenario** (using the concrete details, examples, or contexts provided in the question prompt)

- **MUST incorporate guidance from the retrieved slides** (referencing relevant principles, concepts, frameworks, or information shown in the slide content)

- **MUST make suggestions contextually relevant** to the specific scenario and learning context rather than generic advice

**Question Details:**

- Open-ended question: {Recognize when the multimedia principle has been violated and when it has been applied well. Identify whether there are any part(s) of the lesson that represent a good application and/or violation of the multimedia principle. Provide specific reasons to support your judgment. A teacher is explaining the causes of World War I. On the slide, there's a map of Europe in 1914 and a detailed timeline of alliances forming. At the same time,

an image of a famous WWI painting (a battlefield with soldiers and explosions) slowly fades in for visual impact. The teacher is narrating the complex political tensions and alliances.}

- Retrieved related slides: {as shown in the uploaded version}

- Student's Response: {I think this is a good application because it combines text and images.}

- Model Answer/Key Points: {Good Application: The map of Europe and timeline effectively support learning by providing relevant visual information that helps students understand the complex political relationships and sequence of events leading to WWI. Violation: The battlefield painting represents extraneous visual content that doesn't support the learning objective about causes of the war. This decorative image creates unnecessary cognitive load and distracts from the core instructional content about political tensions and alliances.}

## Task 2: Format Output

After generating the feedback, format your response as a JSON object with this exact structure:

```
{
  "score": "[0 for incorrect, 1 for correct, 2 for partially correct]",
  "feedback": "[A clear, concise revision of the original feedback, retaining key points and removing redundancy. Tooltips are integrated as plain terms. NEVER reveal the correct answer or model answer elements.]",
  "terms": [
    {
      "[term 1]": "[tooltip explanation]"
    },
    {
      "[term 2]": "[tooltip explanation]"
    }
  ],
  "quotes": [
    {
```

```
        "section": "statement",
        "quotes": [
          "[Direct, concise phrases stating correctness or incorrectness.]"
        ]
      },
      {
        "section": "explanation",
        "quotes": [
          "[Phrases explaining why the student's response doesn't fully address th
e question, WITHOUT revealing the correct answer or model answer element
s.]"
        ]
      },
      {
        "section": "advice",
        "quotes": [
          "[Concrete improvement suggestions or reflective prompts.]"
        ]
      }
    ]
}
```

**Formatting Instructions:**

- **CRITICAL**: Maintain the same ABSOLUTE RESTRICTION as Task 1 - DO NOT reveal, mention, hint at, or describe the correct answer, model answer, or any specific elements from the model answer during formatting

- **Terms Extraction**: Identify 2-4 key technical terms or concepts from the feedback that students might need clarification on (e.g., "multimedia principle", "cognitive load", "relevance", "learning objectives", etc.). The terms must use the exact wording as it appears in the feedback text. If no technical terms appear, the terms array can be empty.

- **Quote Extraction Guidelines**:
  - **statement**: Extract the direct judgment phrase(s) about correctness (e.g., "Your answer is incorrect", "partially correct", "correct")

- **explanation**: Extract ALL sentences that explain the student's response and provide reasoning about why it doesn't fully address the question. Include complete sentences and connected ideas, not just fragments. This should capture the full reasoning without revealing correct answers.

- **advice**: Extract sentences that provide concrete suggestions, reflective questions, or learning strategies

- **Complete Sentence Extraction**: When extracting quotes, prioritize complete sentences or meaningful phrase clusters rather than sentence fragments. If multiple sentences form a cohesive explanation, include them together in a single quote.

**Final Output**: Provide only the JSON object in the exact format specified above. No additional explanation, comments, or plain text are allowed.



\end{document}